\documentclass[prl,twocolumn,showpacs,superscriptaddress]{revtex4}
\usepackage{amsmath,amssymb,epsfig}

\begin{document}

\title{Patch coalescence as a mechanism for eukaryotic directional sensing}

\author{A. Gamba}
 \email{andrea.gamba@polito.it}
 \affiliation{Politecnico di Torino and CNISM, Corso Duca
 degli Abruzzi 24, 10121 Torino, Italia}
 \affiliation{INFN, via Pietro Giuria 1, 10125 Torino, Italia}
 \author{I. Kolokolov}
 \affiliation{Landau Institute for Theoretical Physics, Kosygina 2, 119334, Moscow,
 Russia}
 \author{V. Lebedev}
 \affiliation{Landau Institute for Theoretical Physics, Kosygina 2, 119334, Moscow,
 Russia}
 \author{G. Ortenzi}
 \affiliation{Politecnico di Torino and CNISM, Corso Duca degli Abruzzi 24, 10121 Torino,
 Italia}


\begin{abstract}

  Eukaryotic cells possess a sensible chemical compass allowing them to
  orient toward sources of soluble chemicals. The extracellular chemical signal
  triggers separation of the cell membrane into two domains populated by different
  phospholipid molecules and oriented along the signal anisotropy. We propose a
  theory of this polarization process, which is articulated into subsequent stages
  of germ nucleation, patch coarsening and merging into a single domain. We find that
  the polarization time, $t_\epsilon$, depends on the anisotropy degree $\epsilon$
  through the power law $t_\epsilon\propto\epsilon^{-2}$, and that in a cell of
  radius $R$ there should exist a threshold value $\epsilon_\mathrm{th}\propto R^{-1}$
  for the smallest detectable anisotropy.

\end{abstract}

\maketitle

The cells of multicellular organisms are endowed with a chemical compass of amazing
sensitivity, formed as a result of billion years of evolution. Concentration differences
of the order of a few percent in the extracellular soluble attractant chemicals from side
to side are sufficient to induce a chemical polarization of the membrane leading to cell
migration towards the signal source~\cite{SNB+06}. This way, a sensible amplifier of
slight gradients in the distribution of chemicals in the surrounding environment is
realized. Its relevance is easily understood if one recognizes that no multicellular
organism could exist without the constituent cells being able of sensing directional
signals. Directional sensing is actually essential both in embryo development, where
tissue formation is realized through coordinated migration of specific cells guided by
chemical signals, and in the adult organism, where chemical signals guide white blood
cells to the sites of inflammation and platelets to sites of wound repair. The main steps
of the process are as follows (see the reviews~\cite{RSB+03} and~\cite{LH96}). As a
response to the attractant signal, the cell membrane is polarized, afterwards inducing
differentiated polymerization of the cell cytoskeleton in its proximity. The resulting
unbalance, triggered by a well characterized cascade of chemical reactions, leads to the
formation of a growing head and a retracting tail, in such a way that the cell starts to
drift towards the source of the signal. The initial part of this process is constituted
by the early chemical polarization of the cell membrane. In this letter we propose a
simple phenomenological scheme providing a universal description of this fundamental
phenomenon.

Membrane polarization can be recognized as a self-organization process governed by a
network of diffusion-controlled chemical reactions. It is known that reaction-diffusion
networks may become bistable in the presence of chemical feedback
loops~\cite{AS02,AFS04}. In spatially extended systems bistability may lead to the
formation of competing phases and to a phenomenology typical of first order phase
transitions, such as metastability, nucleation and coarsening~\cite{CH93,Bra95}. The
polarized membrane state observed during directional sensing can therefore be interpreted
as the coexistence of domains of two different phases.

Let us briefly describe the chemical reactions which are responsible for directional
sensing. The chemical factors clustering in complementary membrane domains are the
phospholipids PIP$_2$ and PIP$_3$. Two enzymes, PI3K and PTEN, respectively transform
PIP$_2$ into PIP$_3$ and vice\-versa. The phospholipids are permanently bound to the inner
face of the cell membrane, while PI3K and PTEN diffuse in the cell volume and are active
only when they are adsorbed by the membrane. PI3K adsorption takes place through binding
to receptors activated by the extracellular attractant signal. This way, the external
attractant field is coupled to the inner dynamic of the cell. PTEN adsorption takes place
through binding to the PTEN product, PIP$_2$. This process introduces a positive feedback
loop in the system dynamics~\cite{ID02,GCT+05}. When the cell is not stimulated by an
attractant signal the cell membrane is uniformly populated by PTEN and PIP$_2$ molecules.
When a uniform receptor stimulation of a suitable amplitude is switched on, PI3K
molecules bind to the membrane and shift its chemical balance toward a PIP$_3$-rich
phase, while PTEN desorbs. PIP$_3$-rich germs are then nucleated in the PIP$_2$-rich sea
and PIP$_3$-rich regions start to coexist with PIP$_2$-rich ones~\cite{PRG+04}.

Two different regimes of cell polarization may be distinguished. Anisotropy driven
polarization induced by the presence of an attractant gradient is realized in a time of
the order of a few minutes, and results in the formation of a PIP$_3$-rich domain on the
membrane side closer to the attractant source and of a PIP$_2$-rich domain in the
complementary region~\cite{ID02,GCT+05}. On the other hand, cells exposed to uniform
distributions of an attractant polarize in random directions, in times of the order of an
hour (see \emph{e.g.}~\cite{SH85}). The existence of two clearly separated polarization
regimes is confirmed by the recent observation of a sensitivity threshold of the order of
a few percent difference in the attractant molecule concentration from side to
side~\cite{SNB+06}. Direct observation of the polarization process~\cite[Fig. 7a]{ID02}
implies the bound $< 5\,\mathrm{s}$ for PTEN diffusion time in the cell volume, which is
therefore much less than the polarization time for both regimes. In this process, the
amplitude of the cell stimulation is of crucial importance. At very low stimulation
levels PTEN is not desorbed in a significant amount and no directional sensing takes
place. At very high stimulation levels a homogeneous PIP$_3$-rich phase is realized and
directional sensing again does not take place. There exists therefore an optimal
attractant concentration, such that below it the minority phase is PIP$_3$-rich and above
it is PIP$_2$-rich.

Numerical simulations of the directional sensing network performed with the use of
realistic physical and kinetic parameters have shown that under appropriate conditions
the biochemical network is indeed bistable, and that it undergoes spontaneous separation
in chemically different phases, rich in PIP$_2$ and PIP$_3$,
respectively~\cite{GCT+05,CGC+07}. A 5\% anisotropic component in the cell stimulation
accelerates cell polarization and correspondingly decreases the characteristic time
needed for complete phase separation by more than one order of magnitude: fast,
anisotropy driven polarization is realized in times of the order of a minute, while slow,
stochastic polarization is realized in times of the order of one hour, in accordance with
experimentally observed times. In the numerical experiments, when PIP$_2$ is the minority
phase, the evolution leading to phase separation consists of an early nucleation regime,
resulting in the formation of isolated PIP$_2$-rich patches and a late coarsening
process, where large patches of the PIP$_2$-rich phase grow at the expense of the
evaporation of smaller ones, similarly to what happens in the case of first order phase
transitions in a liquid-gas system or in the precipitation of a supersaturated
solution~\cite{LP81}. Finally, the patches condense into a single large cluster, leading
to a stationary state characterized by the coexistence of a PIP$_2$- and a PIP$_3$-rich
domain. However, the dynamics of the directional sensing network differs from that of
otherwise similar processes, such as the precipitation of a supersaturated solution,
under one important respect. When precipitation nuclei in a supersaturated solution
dissolve, matter is transferred to larger nuclei through diffusion in the surrounding
medium. In contrast, in the directional sensing network enzyme-substrate patches
evaporate through desorption of the PTEN enzyme from the membrane, which is then
transferred to other patches through diffusion in the cell volume. Therefore the
transformation of PIP$_3$ into PIP$_2$ molecules cannot be described at the membrane
level as a local, diffusion-like process as is the case with the adsorption-desorption
process from precipitation nuclei; in particular, there is no local conservation of the
number of PIP$_2$ molecules.

\begin{figure}[ht]
 \epsfig{width=7.8cm,file=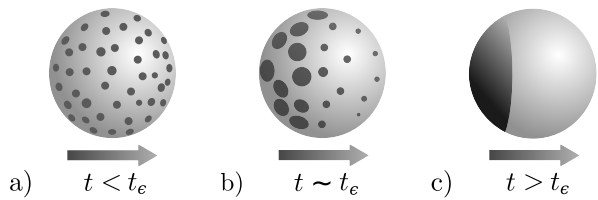} 
  \caption{ Patch growth in the presence of a slight gradient of
  attractant activation directed from left to right.
  The PIP$_3$- and PIP$_2$-rich phases are respectively
  light and dark grey.  }
 \label{fig:uno}
 \end{figure}

The above summarized scenario can be put on a firm analytical ground
resorting to the kinetic theory of first order phase
transitions~\cite{LS61,LP81,Bra95}. In this theory, after germ nucleation
larger patches of the stable phase grow at the expense of smaller patches
which shrink, leading to scaling laws and universal probability
distribution of patch sizes.  We shall now show how the ideas of the
Lifshitz-Slyozov theory~\cite{LS61} may be adapted to our problem to
deduce simple scaling laws for the membrane polarization time and explain
most of the observed phenomenology. We discuss here the case when PIP$_2$
is the minority phase (the other case being symmetric). In this case
PIP$_2$-rich patches are formed inside the PIP$_3$-rich sea (see Fig.
\ref{fig:uno}a). We restrict our consideration to approximately circular
patches of the PIP$_2$-rich phase, which are expected to dominate over
different geometries due to the presence of a linear tension between the
two phases. The free energy of a PIP$_2$-rich patch of radius $a$ can be
written on phenomenological grounds as $F = - \pi\,\psi\, a^2
+2\pi\,\sigma\, a$, where $\sigma$ is the linear tension of the interface
with the surrounding PIP$_3$-rich phase and $\psi$ represents the degree
of metastability~\cite{LP81}, which is a function of the concentration of
PTEN molecules in the cell volume and of the concentration of
extracellular attractant.

According to the kinetic theory of first order phase transitions, the equation of growth
of a patch is dissipative. In the absence of a local conservation law the equation for a
circular patch can be written as $\Gamma\,\partial_t a=-\partial F/\partial a$, where
$\Gamma(a)$ is a damping coefficient~\cite{Bra95}. Since energy dissipation occurs mainly
along the perimeter of the interface between the two phases, $\Gamma$ may be written as
$2\pi a \gamma$, where $\gamma$ is a constant, and we get
 \begin{equation}
 \gamma\,\partial_t a=\psi -\sigma/a +\xi \,,
 \label{gam7}
 \end{equation}
where the noise term $\xi$ represents thermal fluctuations. The fluctuations are
responsible for the formation of an initial population of patches with varying
radii~$a$~\cite[\S 99]{LP81}. Patches with $a$ smaller than the critical radius
$a_\mathrm{c}=\sigma/\psi$ are mainly dissolved while most patches with $a>a_\mathrm{c}$
survive and grow because of the gain in free energy. 
At initial time,
$a_\mathrm{c}$ is of the order of the thickness $a_0$ of the interface between the two
phases~\cite{LP81,Bra95}. As long as the area occupied by patches of the PIP$_2$-rich
phase grows, the degree of metastability $\psi$ decreases, some of the patches that were
initially growing become undercritical and shrink, large patches start ``feeding'' on
smaller ones, and the total number of patches diminishes~\cite[\S 100]{LP81}. In the
final stage of this process a single domain of the PIP$_2$-rich phase is formed
coexisting with the PIP$_3$-rich phase, see Fig.~\ref{fig:uno}c. However, the details of
the process leading to this final stage depend on the external conditions, and,
particularly, on the degree of anisotropy of the attractant signal.

The population of patches can be described in terms of the size distribution function
$f(a)$ such that $f(a) \Delta a$ gives the number of patches with sizes in the interval
$(a, a + \Delta a)$. An important simplification comes from the fact that for patches
with $a>a_\mathrm{c}$ the noise term $\xi$ in (\ref{gam7}) becomes
negligible~\cite{LP81}. This means that the stochastic nature of the problem enters
mainly in the formation of the initial distribution of patch sizes $f(a)$, while for
$a>a_\mathrm{c}$ the time evolution of $f(a)$ is dictated by the deterministic part of
(\ref{gam7}), from which the kinetic equation
 \begin{equation}
  \gamma\, \frac{\partial f}{\partial t}
  + \frac{\partial}{\partial a} \left[
  \left( \psi - \frac{\sigma}{a} \right) f \right] = 0 ,
  \label{gam8}
 \end{equation}
follows~\cite{LP81,Bra95}. With the chosen normalization $\int f(a)\,\mathrm{d}a$
represents the total number of patches, a quantity which is monotonically decreasing in
time according to the previously described ``coarsening'' dynamics~\cite{LS61}. Eq.
(\ref{gam8}) is valid as long as $\int f(a)\,\mathrm{d}a$ is much larger
than~$1$~\cite{LP81}.

To obtain a closed system of equations we need an additional equation for the time
evolution of the metastability degree~$\psi$~\cite{LP81}. In the case of isotropic
stimulation $\psi$ does not depend on the position on the membrane and is instead only a
function of time. Since diffusion of PTEN molecules in the cell volume is faster than
phospholipid diffusion on the membrane we can regard the concentration of PTEN molecules
in the volume as uniform~\cite{GCT+05}. Moreover, fast PTEN diffusion also implies that
$\psi$ instantaneously adjusts to the changes in the size distribution function. While
the total number of patches diminishes as an effect of the coarsening dynamics, the total
area occupied by the patches, as well as the total number of PIP$_2$ molecules found in
the patches, monotonically increases towards their respective equilibrium values. The
metastability degree $\psi$ is equal to zero in equilibrium, and tends to zero in
accordance with
 \begin{equation}
 \psi\propto A-
 \int \mathrm{d} a\ \pi\, a^2 f (t,a) ,
 \label{gam9}
 \end{equation}
as the total patch area tends to its limit value $A$. Eq. (\ref{gam9}) reflects the fact
that in the asymptotic region $\psi$ is proportional to the excess concentration of PTEN
molecules in the volume with respect to the equilibrium value, and therefore to the
difference between the area occupied by the PIP$_2$-rich phase at equilibrium and at
current time. The law (\ref{gam9}) is valid for $t\gg t_0$, where $t_0$ is the
characteristic time needed for the formation of a germ of an alternative phase that can
be estimated as $t_0\sim\gamma\, a_0^2/\sigma$~\cite{LP81,Bra95}.

Asymptotically, (\ref{gam8},\ref{gam9}) lead to the self-similar solution
 \begin{equation}
 \psi(t)=(2\sigma\gamma/t)^{1/2}, \quad
 f \propto t^{-3/2} g\left(\sqrt\gamma\ a/\sqrt{2\sigma t}\right),
 \label{self}
 \end{equation}
where $g(\xi)={\xi}(1-\xi)^{-4} \exp[-{2}(1-\xi)^{-1}]$ if $\xi<1$ and $g(\xi)=0$ if
$\xi>1$. Similarly to what happens in Lifshitz-Slyozov theory~\cite{LS61} the total
number of patches decreases in time due to the evaporation of small patches: $\int
\mathrm{d}a\, f(a)\, \propto t^{-1}$, and from (\ref{self}) one gets $\langle a \rangle=
\sigma/\psi=a_\mathrm{c}$. The evolution of the size distribution function $f$ governed
by (\ref{self}) stops at times of order $t_\star$, where $t_\star$ is defined as the
instant when the average patch size $\langle a \rangle$ reaches the cell size $R$. From
the scaling law $\langle a \rangle\propto\psi^{-1}\propto\sqrt t$ we get $t_\star\sim
(R/a_0)^2\, t_0$. Eventually, at $t\sim t_\star$, a single PIP$_2$-rich patch survives.
Its orientation is determined by the random unbalance in the initial germ distribution.
Notice that in this derivation, following the lines of~\cite{LS61,LP81,Bra95}, isotropy
was essential to assume that $\psi$ was uniform along the whole membrane surface.

Let us now consider the case of an inhomogeneous activation pattern. The inhomogeneity of
the concentration distribution modifies the degree of metastability, which becomes a
function of the position on the membrane surface. Since the distribution of PTEN
molecules in the cell volume is homogeneous, it influences only the isotropic part of the
metastability degree $\psi$, which is a function of time, as previously. In contrast, the
anisotropic part of the metastability degree, $\delta\psi$, related to the external
attractant inhomogeneity, does not depend on time. If the cell membrane has a nearly
spherical form and a radius $R$ much smaller than the characteristic scale of the
extracellular attractant distribution, then $\delta\psi =-\epsilon\, \psi_0\cos\theta$.
Here $\psi_0=\sigma/a_0$ is the initial metastability degree, $\epsilon$ is a
dimensionless factor measuring the initial anisotropy degree, and $\theta$ is the
azimuthal angle on the cell surface. This way we obtain the equation
 \begin{equation}
 \gamma\,\partial_t a=\psi- \epsilon\,\psi_0
 \cos\theta-\sigma/a +\xi,
 \label{eq10}
 \end{equation}
generalizing (\ref{gam7}). As long as $\epsilon\psi_0\ll\psi$, the first stage of patch
growth proceeds approximately as in the isotropic case and $\psi$ decreases as
$t^{-1/2}$. However, at a time of order $t_{\epsilon}$, where $t_{\epsilon}$ is defined
by the equation $\psi (t_{\epsilon}) = \epsilon\,\psi_0$, the perturbation
$\epsilon\,\psi_0\cos\,\theta$ becomes comparable to $\psi$ and the process of
polarization becomes anisotropic, so that patches in different regions get different
average sizes, see Fig.~\ref{fig:uno}b. From the scaling law (\ref{self}) for $\psi$ one
gets $t_\epsilon\sim t_0\,\epsilon^{-2}$. For $t>t_\epsilon$, the leading term in
(\ref{eq10}) becomes the perturbation $\epsilon\,\psi_0\cos\theta$, implying that in the
region closer to the source of the stimulation ($\cos\,\theta\gtrsim 0$) the PIP$_2$-rich
phase evaporates in a time which is easily estimated as being again of order
$t_\epsilon$, leading to the formation of a single PIP$_2$-rich patch in the region
further from the source of the stimulation ($\cos\,\theta\lesssim 0$) and realizing
complete polarization, as shown in Fig.~\ref{fig:uno}c.

The above scheme is valid as soon as the initial nucleation time $t_0$ is significantly
smaller than $t_{\epsilon}$, an assumption which is compatible with the results of
numerical experiments~\cite{GCT+05}. On the other hand, the second stage of patch
evolution occurs only if $t_\star\gg t_\epsilon$. Otherwise, the presence of a gradient
of attractant becomes irrelevant and only the stage of isotropic patch growth actually
occurs. This condition implies that a smallest detectable gradient exists, such that
directional sensing is impossible below it. The threshold value $\epsilon_\mathrm{th}$
for $\epsilon$ is found by letting $t_\star=t_\epsilon$. Since the product
$\psi\,a_\mathrm{c}$ is a time-independent constant, we can simply compare its value at
initial and final time when $\epsilon= \epsilon_\mathrm{th}$, obtaining
$\epsilon_\mathrm{th}={a_0}/{R}$, which gives us the expression for the threshold
anisotropy.

It is interesting to estimate $a_0$, and, consequently, $\epsilon_\mathrm{th}$, in terms
of observable parameters. Comparing the characteristic patch surface and perimeter energy
as a function of the phospholipid diffusion coefficient $D$, surface phospholipid
concentration $c$, surface concentration of activated receptors $h$, and the
characteristic catalytic time $\tau$, one gets $a_0\sim(D\,\tau\,c/h)^{1/2}$. Using
parameter values from Ref.~\cite{GCT+05} one gets $a_0\sim 1\,\mu\mathrm{m}$ and
$\epsilon_\mathrm{th}\sim 10\% $. The value for $\epsilon_\mathrm{th}$ is compatible with
the observations (the data from Ref.~\cite{SNB+06} imply $\epsilon_\mathrm{th}\simeq 7\%$
for \textit{dictyostelium}).

One may wonder whether a cell may become polarized by the anisotropy produced by a
spontaneous fluctuation in the extracellular distribution of attractant molecules or
fluctuations in receptor-ligand binding~\cite{LH96}. Since eukaryotic cells typically
carry $10^4$-$10^5$ receptors for attractant factors, one expects spontaneous
fluctuations in the fraction of activated receptors to be of the order of $10^{-2}$, a
value which is comparable to observed anisotropy thresholds. However, to actually produce
directed polarization the fluctuation should sustain itself for several minutes,
\textit{i.e.} for a time comparable to the characteristic polarization time. Such an
event has very low probability of being observed since the correlation time of the
fluctuations determined by attractant diffusion at the cell scale and the characteristic
times of receptor-ligand kinetics are much less than the polarization time. Indeed, the
diffusion time is $\sim 1\,\mathrm{s}$ at the typical cell size $10\,\mu\mathrm{m}$, and
the characteristic times of receptor-ligand kinetics are also $\sim 1\,\mathrm{s}$ (see
online supporting information to Ref.~\cite{SNB+06}). Therefore, the direction of cell
polarization in the case of a homogeneous distribution of attractant can only be
determined by the inhomogeneity in the initial distribution of the positions of
PIP$_2$-rich germs produced by thermal fluctuations.

In conclusion, we have constructed a universal phenomenological description of the
mechanism of directional sensing in the eukaryotes based on the process of patch
coarsening. This description implies the existence of two clearly separated polarization
regimes depending on the presence or absence of an anisotropic component in the
activation pattern produced by the extracellular attractant factor, and the existence of
a sensitivity threshold for the anisotropic component. Both results are in reasonable
agreement with experimental observations. Moreover, we predict that directed polarization
time should scale as the inverse square of the relative signal anisotropy, a law that
should be verifiable by direct observation. Our picture suggests that directed and
stochastic polarization share a common mechanism, and that stochastic polarization should
be the result of noise in subcellular and not in extracellular dynamics. Importantly, our
picture does not depend on the details of the reactions involved, but only on the general
structure of the directional sensing network and on its bistability. This means that the
picture is robust not only with respect to variations of the kinetic and physical
parameters, but also with respect to the identity of the chemical species involved.
Indeed, PI3K and PTEN could be substituted by, or synergize with, molecules endowed with
similar enzymatic activity. An interesting speculation is that the bound
$\epsilon_\mathrm{th}={a_0}/{R}$ may explain why spatial directional sensing was
developed only in the large eukaryotic cells and not in smaller prokaryotes, whose
directional sensing mechanisms rely instead on the measurement of temporal variations in
concentration gradients~\cite{ASB+99}. Our bound derives from the intrinsic properties of
polarization dynamics and is independent of the size criterion formulated in
Ref.~\cite{BP77}. The experimental observation of selforganized phospholipid
patches~\cite{PRG+04} 
following uniform attractant stimulation provides an initial confirmation of the validity
of our scheme. To check the predictions of our theory, similar observations should be
performed for the longer times characteristic of random and directed polarization, both
under uniform attractant activation and in the presence of accurately controlled
concentration gradients. Experimental modulation of PTEN levels could be used to modify
the overall size of patches and eventually switch off the patch formation mechanism.

 \begin{acknowledgments}
A.G. likes to thank Guido Serini and Stefano Di Talia for many helpful discussions. I.K.
and V.L. acknowledge partial support of RFBR grant 06-02-17408-a.
 \end{acknowledgments}

\end{document}